\documentclass[useAMS,usenatbib]{mn2e} 
\usepackage{amsmath,amssymb}
\usepackage{graphicx} 
\usepackage{aas_macros}
\usepackage[usenames]{color} 
\usepackage{sidecap}

\title[Spots on Am stars]{Spots on Am stars}
\author[L. A. Balona, G. Catanzaro, O. P. Abedigamba, V. Ripepi, B. Smalley]{L. A. Balona$^1$,
G. Catanzaro$^2$, O. P. Abedigamba$^3$, V. Ripepi$^4$, B. Smalley$^5$\\
\\
$^1$South African Astronomical Observatory, P.O. Box 9, Observatory 7935, Cape Town, South
Africa\\
$^2$INAF-Osservatorio Astrofisico di Catania, Via S.Sofia 78, I-95123, Catania, Italy\\ 
$^3$Department of Physics, North-West University, Private Bag X2046, Mmabatho, 2735, South Africa\\
$^4$INAF-Osservatorio Astronomico di Capodimonte, Via Moiariello 16, I-80131, Napoli, Italy\\ 
$^5$Astrophysics Group, Keele University, Keele, Staffordshire, ST5 5BG, UK
}

\date{} 
 
\pagerange{\pageref{firstpage}--\pageref{lastpage}} \pubyear{2002} 
 
\def\LaTeX{L\kern-.36em\raise.3ex\hbox{a}\kern-.15em 
 T\kern-.1667em\lower.7ex\hbox{E}\kern-.125emX} 
 
\begin{document} 
 
\label{firstpage} 
 
\maketitle 
 
\begin{abstract} 
We investigate the light variations of 15 Am stars using four years of
high-precision photometry from the {\it Kepler} spacecraft and an additional
14 Am stars from the K2 Campaign 0 field.  We find that most of the Am stars
in the {\it Kepler} field have light curves characteristic of rotational
modulation due to star spots.  Of the 29 Am stars observed, 12 are
$\delta$~Scuti variables and one is a $\gamma$~Doradus star. One star is an
eclipsing binary and another was found to be a binary from time-delay
measurements.  Two Am stars show evidence for flares which are unlikely to be 
due to a cool companion.  The fact that 10 out of 29 Am stars are rotational variables and
that some may even flare strongly suggests that Am stars possess significant
magnetic fields.  This is contrary to the current understanding that the
enhanced metallicity in these stars is due to diffusion in the absence of a
magnetic field.  The fact that so many stars are $\delta$~Scuti variables is
also at odds with the prediction of diffusion theory.  We suggest that a
viable alternative is that the metal enhancement could arise from accretion.
\end{abstract} 


\section{Introduction}

The Am stars (metallic-lined A stars) are a group of A- and early F-type stars 
in which the CaII K line is relatively weak, and/or metallic lines relatively 
strong, compared with the spectral type indicated by the Balmer lines. 
More detailed analysis shows that Am stars are characterized by an under-abundance 
of Ca (and/or Sc) and/or an over-abundance of Fe and the Fe-group elements.  
The anomalous abundances in Am and Ap stars are thought to be a result of
the interplay between gravitational settling and radiative acceleration of
different atomic species.  As a result, some elements diffuse upwards and
others settle downwards provided there is little mixing.  Mixing due to
convection or meridional circulation due to rotation destroys the natural
segregation of elements due to diffusion that would otherwise occur. 
Convection in A stars is confined to a very thin sub-surface layer which means
that the diffusion process should proceed unhindered in slowly-rotating A
stars.  

The difference between Ap and Am stars is attributed to the fact that Ap stars 
have strong magnetic fields, whereas Am stars do not.  \citet{Auriere2010} 
have confirmed that there is at least an order of magnitude difference in 
magnetic field strength between Ap/Bp stars and Am and HgMn stars.  The 
presence of a large-scale strong dipole magnetic field in Ap stars modifies 
the diffusion rate in accordance with the geometry of the field and is 
responsible for the abundance patches in Ap stars.  On the other hand, the 
weak (or perhaps tangled) magnetic fields in Am stars leads to a relatively 
homogeneous abundance distribution across the stellar surface.

Early calculations predict a much larger over-abundance of heavy elements 
in Am stars than actually observed.  \cite{Richer2000} found that it is 
important to consider the diffusion of heavy elements as well as helium.
As the heavy elements settle, enrichment of the deeper iron bump opacity 
region of the iron-group causes a convective zone to develop in that region 
which mixes with the convective envelope by overshooting.  The result is a 
dilution of the abundance anomalies, leading to abundances more in 
line with what is observed.  More recent work shows that it is also important 
to include the destabilizing effect (thermohaline instability) caused by the
composition gradient produced by diffusion ($\mu$ gradient) \citep{Vauclair2012}.

It is thought that stars in binaries with orbital periods of 2--10\,d become 
Am stars because their rotational velocities, $v$, have been reduced by tidal 
interactions below $v \approx 120$\,km\,s$^{-1}$, a requirement for diffusion 
to act. Both Am and normal stars occur in binaries with orbital periods between
10--100\,d, but why this is the case is not understood.  A recent search for
eclipsing binaries among Am stars suggests that around 60--70\,percent of Am
stars are spectroscopic binaries \citep{Smalley2014}, which is consistent
with radial velocity studies \citep{Abt1985, Carquillat2007}.  Clearly, a
substantial number of Am stars are probably single or in wide binaries and
it is no longer clear whether binarity is a requirement for the Am phenomenon.

The effect of diffusion on pulsation of Am stars is to reduce the width of the 
instability strip, with the blue edge shifting towards the red edge, eventually 
leading to the disappearance of instability when helium is sufficiently depleted 
from the HeII ionization zone \citep{Turcotte2000}. The amount of driving is 
related to the amount of He still present in the driving zone.  The cooler the 
star, the deeper the mixing and the greater the He abundance in the driving 
zone.  Models therefore predict that Am stars should not pulsate unless they
are sufficiently cool.  More recent calculations suggest that rotational 
mixing is important in maintaining the He abundance in the driving region 
\citep{Talon2006}.

These ideas can be tested if we can obtain sufficiently accurate physical
parameters to place pulsating and non-pulsating Am stars in the H-R diagram.
\citet{Smalley2011} studied over 1600 Am stars and found that around 200 Am 
stars are pulsating $\delta$~Sct and $\gamma$~Dor stars.  It seems that
Am stars, in general, seem to pulsate at somewhat lower amplitudes
than normal $\delta$~Sct stars.  However, \citet{Smalley2011} could find
no real difference between the location of pulsating Am stars and $\delta$~Sct 
stars in the instability strip, except for a slight tendency of pulsating 
Am stars to be cooler than normal $\delta$~Sct stars.  \citet{Catanzaro2012} 
also concluded that there is a tendency for pulsating Am stars to be cooler than
normal $\delta$~Sct stars and that the percentage of pulsating Am stars is 
about the same as the percentage of $\delta$\,Sct stars in the instability
strip.  Similar conclusions were obtained by \citet{Balona2011d}.  It is 
clear that the prediction of diffusion theory regarding pulsation needs to 
be re-examined.  The diffusion process, in fact, seems to play only a minor 
role in determining the pulsational stability of A and early F stars.  Even 
Ap stars pulsate as $\delta$~Sct stars at low amplitudes \citep{Balona2011a}.  

Recent high-precision photometric observations of A stars using {\it Kepler} 
have further undermined the general consensus that these stars have stable
envelopes where diffusion has a strong competitive advantage.  It seems that 
about 2\,percent of A stars observed by {\it Kepler} show flares
\citep{Balona2012c, Balona2014b}.  It can be easily demonstrated that the
flares are associated with the A star and not with a presumed cool companion.   
Furthermore, the {\it Kepler} data suggest that at least 40 percent of A 
stars are rotational variables \citep{Balona2013c}, presumably due to starspots.  
Since it is generally believed that Am stars do not have magnetic fields, 
the incidence of rotational modulation should be absent or small in Am stars 
compared to normal A stars.  

The previous study of {\it Kepler} Am stars \citep{Balona2011d} used only 
data for approximately the first 50\,d and was mainly aimed at detecting 
$\delta$~Sct pulsations in these stars.  The low-amplitude light variations 
due to rotational modulation were suspected but could not be studied.  There
are now four years of almost continuous high-precision photometry for these 
stars, making it easier to detect rotational modulation.   In this paper we 
investigate rotational modulation in Am stars using {\it Kepler} data. 
Our aim is to establish whether the incidence of rotational modulation or other 
activity is lower in Am stars than in normal A stars.  The answer might explain 
why there are many slowly-rotating A stars which are not Am stars and provide 
further clues as to the nature of the Am phenomenon.

\section{The data}

 There are 15 known Am stars in the {\it Kepler} field  \citep{Balona2011d}.  
We also include 14 Am stars observed by the {\it Kepler} K2 Campaign 0 mission.  
These data are far less favourable to detecting rotational modulation because 
only about 45\,d of data are available and the photometric precision is 
severely compromised by the telescope motion.  However, The K2 data are
useful in detecting $\delta$~Sct pulsations in these stars.

For the vast majority of stars, {\it Kepler} photometry is available only
in long-cadence (LC) mode which consist of practically uninterrupted 30-min
exposures covering a time span of over three years.  For a few thousand stars, 
including all the Am stars discussed here, short-cadence (SC) 1-min exposures 
are also available, but these usually cover only one or two months.
Characteristics of LC data are described in \citet{Jenkins2010a} 
while \citet{Gilliland2010a} describe the characteristics of SC data.   The 
science processing pipeline is described by \citet{Jenkins2010b}.  The 
{\it Kepler} data are divided into quarters of approximately 90-d duration 
which are numbered Q0, Q1, Q2, etc., Q0 being the initial 10-d commissioning run.  
In this study we use all available LC data  (Q0 -- Q17) from JD~2454953 to
JD~2456424 (1471~d).  These data are publicly available on the Barbara A. 
Mikulski Archive for Space Telescopes (MAST, {\tt archive.stsci.edu}). 

The light curve files contain simple aperture photometry (SAP) values and a 
more processed version of SAP with artifact mitigation included called 
Pre-search Data Conditioning (PDC) flux.  The SAP photometry contains 
discontinuities, outliers, systematic trends and other instrumental 
signatures.  The PDC tries to remove these errors while preserving 
astrophysically interesting signals by using a subset of highly correlated and
quiet stars to generate a cotrending basis vector set.  The PDC flux uses the 
best fit that simultaneously removes systematic effects while reducing the 
signal distortion and noise injection.  Details of how this is accomplished
are described in \citet{Stumpe2012} and \citet{Smith2012}.  Because we
are interested in low-frequency variations, we have used the PDC
fluxes in this analysis.

The {\it Kepler} SAP data contains drifts and jumps.  There is a zero-point 
difference between data in different quarters and the instrumental drift 
varies from quarter to quarter.  Although the PDC data are designed to
remove these effects, it is possible that residual long-term
instrumental effects are still present at a low level.  To understand how
correction of data may effect the low-frequency end of the periodogram, we
performed a simple simulation where we applied our own correction to the SAP 
data of a few stars which appeared to be the least variable among 20000
{\it Kepler} stars.  To correct the SAP data we fitted and removed polynomials
of low degree (typically degree 3 or 5) to individual quarters. The polynomials
do not join smoothly between any two quarters, but we ignored these small 
jumps.  To these data we added sinusoidal variations comprising of 100 
frequencies of equal amplitudes and random phases. These frequencies start at 
zero frequency and are equally spaced by $0.1$\,d$^{-1}$.  The periodogram 
recovers all frequencies, including the lowest frequency of 0.1\,d$^{-1}$.  
The same simulation using a signal of 0.02\,d$^{-1}$ recovers a lowest 
frequency of 0.08\,d$^{-1}$.  From this simulation, we may deduce that 
frequencies higher than about 0.08\,d$^{-1}$ are not affected by the crude 
correction procedure.  The corrections used to obtain the PDC data are, of 
course, more sophisticated and we can safely assume that frequencies higher 
than 0.08\,d$^{-1}$ are easily recovered.

The light curves from the extended {\it Kepler} K2 Campaign 0 mission were 
obtained from the raw pixel data files using a suitable aperture.  Our own
software, {\tt keplc}, was used for this purpose.  For fields where
there were few or faint stars in close proximity to the target, we used the
largest possible aperture in order to minimize the instrumental light
variation due to image motion.  The telescope motion can be found by tracking
the centroid of the star.  There is a 4.08\,d$^{-1}$ periodic component in
the image motion which is reflected in the light curve.  We therefore
removed this frequency and all its harmonics from the light curve so as to
enhance the actual stellar signal.  Long-term drifts and jumps were also
removed.  The periodogram of the processed light curve was then calculated.
Due to image motion and long-term drifts, the periodogram noise level for
the K2 data increases strongly towards zero frequency.  As a result, it is
generally not possible to detect low frequencies in the star unless these
are of high amplitude.

\begin{table*}
\caption{List of Am stars in the {\it Kepler} field (top) and the K2 Campaign 
0 field (bottom).  The stars are identified by their KIC number and their
EPIC number respectively while the second column is the HD number.  The spectral 
classification is given in the third column.  The apparent magnitude and the 
derived effective temperature, $T_{\rm eff}$, are mostly from the KIC except 
where noted.  The surface gravity, $\log g$ and projected rotation velocity, 
$v \sin i$, are from \citet{Uytterhoeven2011} except where noted.  The variable 
star classification is estimated from the light curve and periodogram.  The 
rotation period of the  star ($P_{\rm rot}$ in days) and the rotational light 
amplitude ($A$ in ppm) is estimated from the periodogram.  References are 
as follows:  1 - \citet{Abt1981}, 2 - \citet{Abt1984b},  3 - \citet{Ammons2006}, 
4 - \citet{Antoci2011}, 5 - \citet{Bertaud1960}, 6 - \citet{Catanzaro2011}, 
7 - \citet{Catanzaro2014}, 8 - \citet{Catanzaro2015}, 9 - \citet{Floquet1970}, 
10 - \citet{Floquet1975}, 11 - \citep{Halbedel1985}, 12 - \citet{McCuskey1967}, 
13 - \citet{McCuskey1959}, 14 - \citet{Renson2009}.}
\label{amtab}
\begin{tabular}{rrlrllllrr}
\hline
\multicolumn{1}{c}{KIC/EPIC} &
\multicolumn{1}{c}{HD} &
\multicolumn{1}{c}{Sp. Type} & 
\multicolumn{1}{c}{V} &
\multicolumn{1}{c}{$T_{\rm eff}$} &
\multicolumn{1}{c}{$\log g$} &
\multicolumn{1}{c}{$v \sin i$} &
\multicolumn{1}{c}{Type} &
\multicolumn{1}{c}{$P_{\rm rot}$} &
\multicolumn{1}{c}{$A$} \\
\multicolumn{1}{c}{} &
\multicolumn{1}{c}{} &
\multicolumn{1}{c}{} & 
\multicolumn{1}{c}{} &
\multicolumn{1}{c}{(K)} &
\multicolumn{1}{c}{(dex)} &
\multicolumn{1}{c}{(km\,s$^{-1}$)} &
\multicolumn{1}{c}{} &
\multicolumn{1}{c}{(d)} &
\multicolumn{1}{c}{(ppm)} \\
\hline
  3429637  & 178875  & kF2hA9mF3$^2$     &  7.71 &  7100$^{6}$ & 4.00       & ~50         & DSCT      &         &      \\
  3836439  & 178661  & kA2.5hF0mF0$^1$   &  7.57 &  7166       & 3.90       &             & EA        &         &      \\
  5219533  & 226766  & kA2hA8mA8$^2$     &  9.20 &  7409       & 3.90       & 115         & DSCT      &         &      \\  
  7548479  & 187547  & A4Vm$^4$          &  8.40 &  7500$^4$   & 3.90       & ~10         & DSCT/ROT  &  0.6530 &  150 \\
  8323104  & 188911  & A2mF0$^{10}$      &  9.66 &  7587       & 3.90       &             & ROT       &  6.0447 &    7 \\
  8703413  & 187254  & A2mF0$^{10}$      &  8.71 &  7960$^{7}$ & 3.80$^{7}$ & ~15$^{7}$   & ROT       &  6.5292 &  215 \\
  8881697  &         & A5m$^5$           & 10.57 &  7732       & 3.90       &             & DSCT/ROT  &  0.4319 &  200 \\
  9117875  & 190165  & A2mF2$^{10}$      &  7.51 &  7460$^{7}$ & 3.60$^{7}$ & ~58$^{7}$   & ROT       &  1.4949 &   42 \\
  9204718  & 176843  & A3mF0$^{10}$      &  8.79 &  7610$^{7}$ & 3.80$^{7}$ & ~27$^{7}$   & DSCT/ROT  &  8.7308 &   90 \\
  9272082  & 179458  & A5m$^9$           &  9.00 &  8150$^{7}$ & 3.90$^{7}$ & ~75$^{7}$   & ROT       &  0.9703 &    4 \\
  9349245  & 185658  & Am:$^{8}$         &  8.15 &  8300$^{8}$ & 4.00$^{8}$ & ~80$^{8}$   & ROT       &  0.9420 &    7 \\
  9764965  & 181206  & Am:$^{8}$         &  8.85 &  7800$^{8}$ & 3.88$^{8}$ & ~87$^{8}$   & DSCT/ROT  &  0.4872 &  135 \\
 11402951  & 183489  & kF3hA9mF5$^2$     &  8.13 &  7150$^{6}$ & 3.50       & 100         & DSCT      &         &      \\
 11445913  & 178327  & kA7hA9mF5$^2$     &  8.49 &  7200$^{6}$ & 3.50       & ~51         & DSCT      &         &      \\
 12253106  & 180347  & Am$^{8}$          & 8.41  &  7900$^{8}$ & 3.85$^{8}$ & ~12$^{8}$   & ROT       &  4.1043 &   82 \\
\\
202059291  & 40788   & A2$^{12}$, A2m$^{14}$ &  9.04 &  7576$^{3}$ &             &            & SB2       &  2.984  &  188 \\
202059336  & 43509   & Am$^{8}$          &  8.90 &  7900$^{8}$ & 3.97$^{8}$  & ~28$^{8}$  &           &         &      \\
202061329  & 250408  & A2$^{12}$, A2m$^{14}$   & 10.36 &  8329$^{3}$ &             &            &           &         &      \\
202061333  & 250786  & A7$^{12}$, A5m$^{14}$   & 10.40 &  7431$^{3}$ &             &            & DSCT      &         &      \\
202061336  & 251071  & F0V$^{12}$, A3m$^{14}$  & 10.34 &  7316$^{3}$ &             &            &           &         &      \\
202061338  & 251150  & A2$^{12}$, A0m$^{14}$   & 10.57 &  7959$^{3}$ &             &            & DSCT      &         &      \\
202061349  & 251947  & A3p:$^{12}$, A2m$^{14}$ & 10.95 &  6561$^{3}$ &             &            &           &         &      \\ 
202061353  & 252154  & F2V$^{12}$, A5m$^{14}$  &  9.85 &  6925$^{3}$ &             &            & DSCT      &         &      \\
202061363  &         & A0m..$^{13}$    & 11.00 &  4569$^{3}$ &             &            &           &         &      \\
202062436  &         & A2-F0$^{12}$    & 11.64 &  6500       &             &            & DSCT      &         &      \\
202062447  & 251717  & A3-F2$^{12}$    & 11.20 &  7396$^{3}$ &             &            & GDOR      &         &      \\
202062450  & 256320B & kA5hA8mF5$^{11}$  & 10.80 &             &             &            &           &         &      \\
202062454  & 251812  & A5-F2:$^{12}$   & 11.61 &  6750$^{3}$ &             &            &           &         &      \\
202062455  & 251463  & A5-F2$^{12}$    & 10.70 &  6220$^{3}$ &             &            &           &         &      \\  
\hline                                                                                                
\end{tabular}
\end{table*}

\section{{\it Kepler} light curves}

\begin{figure}
\centering
\includegraphics[]{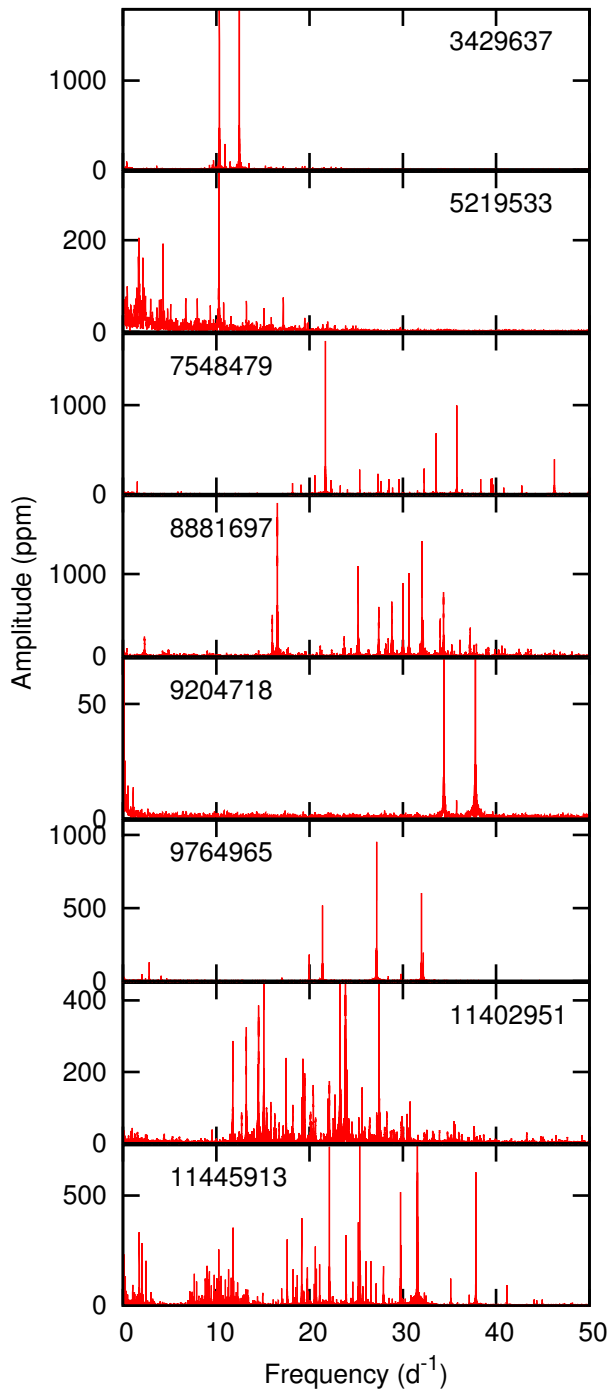} 
\caption{Periodograms of short-cadence light curves of Am
stars in the {\it Kepler} field which are $\delta$~Scuti variables.}
\label{dsctsc} 
\end{figure}

\begin{figure}
\centering
\includegraphics[]{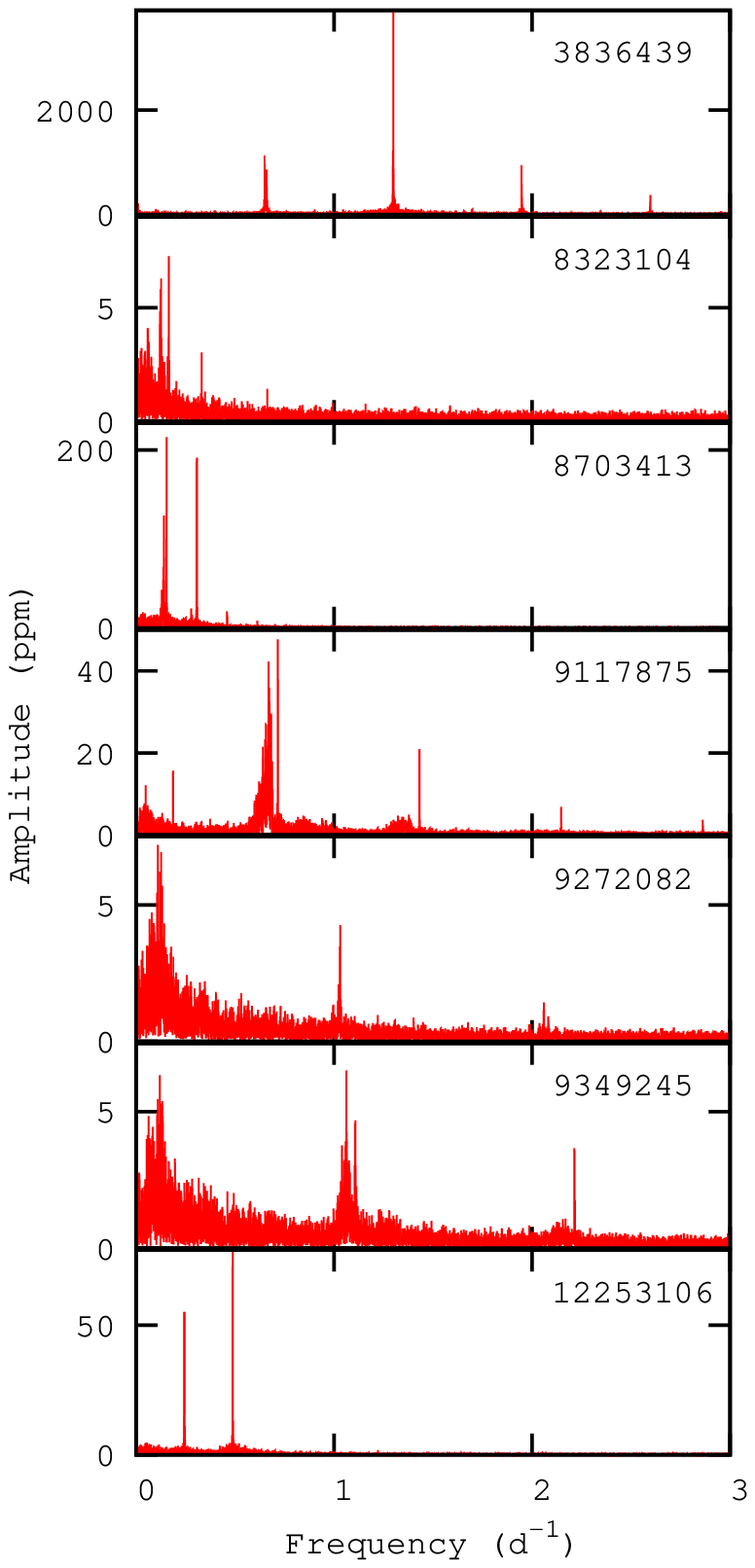} 
\caption{Periodograms of long-cadence light curves of Am
stars in the {\it Kepler} field which are not $\delta$~Sct 
variables.}
\label{nodsct} 
\end{figure} 

A list of known Am stars in the {\it Kepler} and K2 Campaign 0 fields is 
shown in Table\,\ref{amtab}.  The identifications are from the {\it Kepler}
Input Catalogue (KIC, \citet{Brown2011a}) and from the K2 Ecliptic Plane
Input Catalog (EPIC, {\scriptsize \tt https://archive.stsci.edu/k2/KSCI-19082-002.pdf}).
The spectral classification for Am stars in the {\it Kepler} field are mostly 
quite recent and detailed metal abundances are available for most of the
stars.  The Am nature of these stars is hardly in doubt.  The spectral 
classification for Am stars in the K2 field are far less secure.  All of them 
are listed as Am stars by \citet{Renson2009}, but the original reference is 
often missing.  In Table\,\ref{amtab} we have referenced what we believe to 
be the original source, but sometimes the Am nature is not even mentioned.
Part of the problem may be in the way the recognition of the Am class has 
evolved with time. 

Examination of the light curves shows that of the 29 Am stars, there are 12 
$\delta$~Sct variables, 1 $\gamma$~Dor variable and 2 detached eclipsing 
binaries.  Most of the others have light curves similar to those generally 
attributed to rotational modulation caused by starspots. Periodograms of the 
$\delta$~Sct Am stars in the {\it Kepler} field are shown in 
Fig.\,\ref{dsctsc}.  Periodograms of the other stars in the {\it Kepler}
field are shown in Fig.\,\ref{nodsct}. 

It is evident from Fig.\,\ref{nodsct} that highly significant, coherent, 
low-frequency variations and harmonics are present in all six non-$\delta$~Sct
stars.  In KIC\,3836439 this can be attributed to proximity effects in a
binary.  \citet{Luyten1936} found the star to be a spectroscopic binary with 
$P_{\rm orb} = 1.54039$\,d, semi-amplitude $K = 88.7 \pm 2.8$\,km\,s$^{-1}$ 
and a circular orbit.  \citet{Batten1978} further lists the mass function 
$f(m) = 0.112$ and $a_1 = 1.88\times 10^6$\,km $= 2.68 R_\odot$.  {\it Kepler}
observations confirm that this is an eclipsing binary with $P = 1.540407$\,d 
in a detached Algol-type system with $\sin i = 0.96$ \citep{Slawson2011}.  
The ratio of effective temperatures $T_2/T_1 = 0.758$ and the sum of radii 
$R_1+R_2 = 0.35a$ where $a$ is the semi-major axis.  Slightly different results
are given by \citet{Prsa2011}: $\sin i = 0.95$, $T_2/T_1 = 0.48$ and 
$R_1+R_2 = 0.37a$.  It is possible that the Am classification is a result of a
composite spectrum.

\begin{figure}
\centering
\includegraphics[]{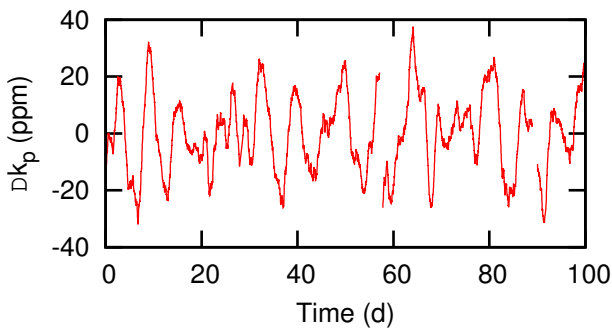} 
\caption{Portion of the corrected {\it Kepler} light curve of KIC\,8323104
showing regular variations.}
\label{k008323104} 
\end{figure}

In the other stars, the low-frequency variations are unlikely to be due to
proximity effects in a binary because the amplitudes often vary and the
periodogram lines attributed to rotation are broadened in some stars.  This is to be
expected in a rotational variable because spots change in size and intensity
with time.  Differential rotation, which appears to be a general feature in
A stars \citep{Balona2013c}, will also cause such broadening.  As an example
of rotational modulation, Fig.\,\ref{k008323104} shows part of the 
long-cadence PDC light curve of KIC\,8323104.  Quasi-regular light variability
is present with a peak-to-peak amplitude of about 40\,ppm and a timescale of 
around 6\,d.  This cannot be a result of proximity effects in a binary 
because of the irregularity of the variations.  The most plausible explanation 
is that it is a result of rotational modulation due to stellar activity.

The second-last column of Table\,\ref{amtab} shows the rotation period estimated
from the periodogram.  In the case of $\delta$~Sct stars there are often
many other low-frequency peaks which are always present in any $\delta$~Sct
star.  In this case we identified the rotation peak because of its
relatively high amplitude and the presence of an harmonic \citep{Balona2013c}.  
The presumed rotation period decreases with $v \sin i$, as might be expected.

\begin{figure} 
\centering
\includegraphics[]{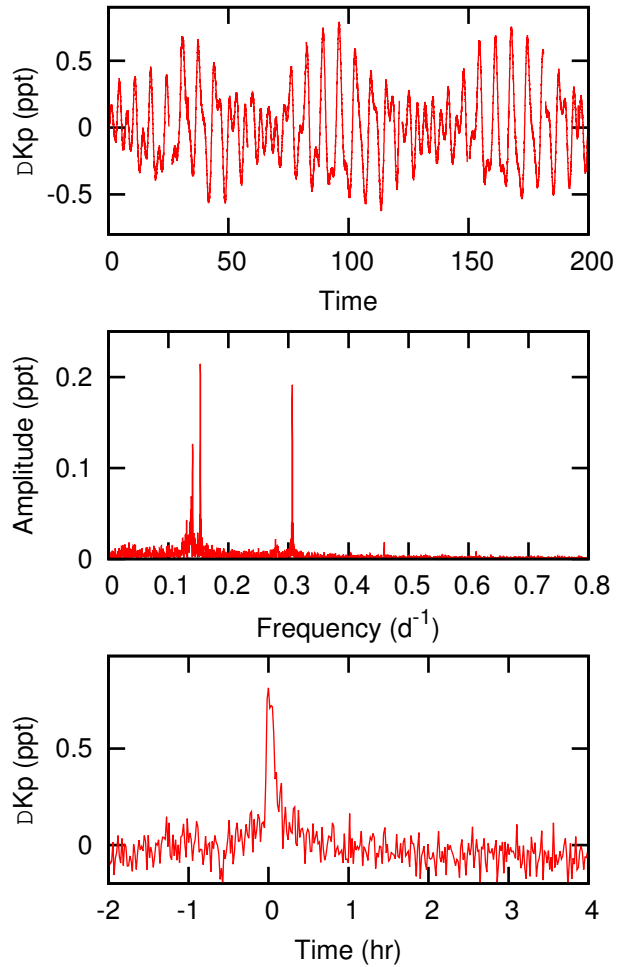} 
\caption{Top panel: part of the {\it Kepler} long-cadence light curve of
KIC\,8703413 showing traveling features typical of starspots.  Middle pane:
periodogram of all LC data.  Bottom panel: presumed flare at JD\,2455164.88
seen in SC data.}
\label{k008703413}
\end{figure}

KIC\,8703413 was first reported as an Am star by \citet{Mendoza1974} and
subsequently classified as kA2mF0 by \citet{Floquet1975}. The star was
confirmed as Am by \citet{Catanzaro2014}.  As can be seen in
Fig.\,\ref{k008703413}, the {\it Kepler} light curve shows distinct beating 
and traveling wave characteristic of star spots \citep{Uytterhoeven2011,
Balona2013c}.  The periodogram shown in the middle panel has a main peak at 
$f_1 = 0.1532$\,d$^{-1}$ and its harmonics.  In addition, there is another 
peak $f_2 = 0.1404$\,d$^{-1}$ which looks to be a bit broader.  It is not 
clear how this pattern may be interpreted, but it seems reasonable to suppose 
that either $f_1$ or $f_2$ could be the rotation frequency.   Perhaps both 
frequencies could be ascribed to starspots if the star is in differential 
rotation.  No other frequencies are present.  The projected rotational 
velocity, $v \sin i = 15 \pm 2$\,km\,s$^{-1}$, is compatible with the rotation
frequency, giving a radius $R \sin i\approx 1.96 R_\odot$.  There are seven 
radial velocities in the literature \citep{Fehrenbach1997}, which is not 
enough for an analysis.  The velocity range is quite large and it is
possible that the star may be a binary.  

KIC\,8703413 is a flare star \citep{Balona2012c} as can be seen in the bottom 
panel of Fig.\,\ref{k008703413}.  The flare intensity is about 0.7\,ppt. The 
A star has a luminosity of about 100 times larger than a K dwarf,  which means 
that if the flare is attributed to an unseen K dwarf companion, its relative 
intensity would have to be about 70\,ppt.  The mean flare intensities in
{\it Kepler} field K-dwarfs are about 3.3\,ppt \citep{Walkowicz2011}, which 
means that this particular flare would have to be of unusual intensity.  In 
general, flares in A stars cannot be attributed to a cool companion since the 
average flare intensity is about two orders of magnitude larger than that in 
a typical K or M dwarf.

\begin{figure}
\centering
\includegraphics[]{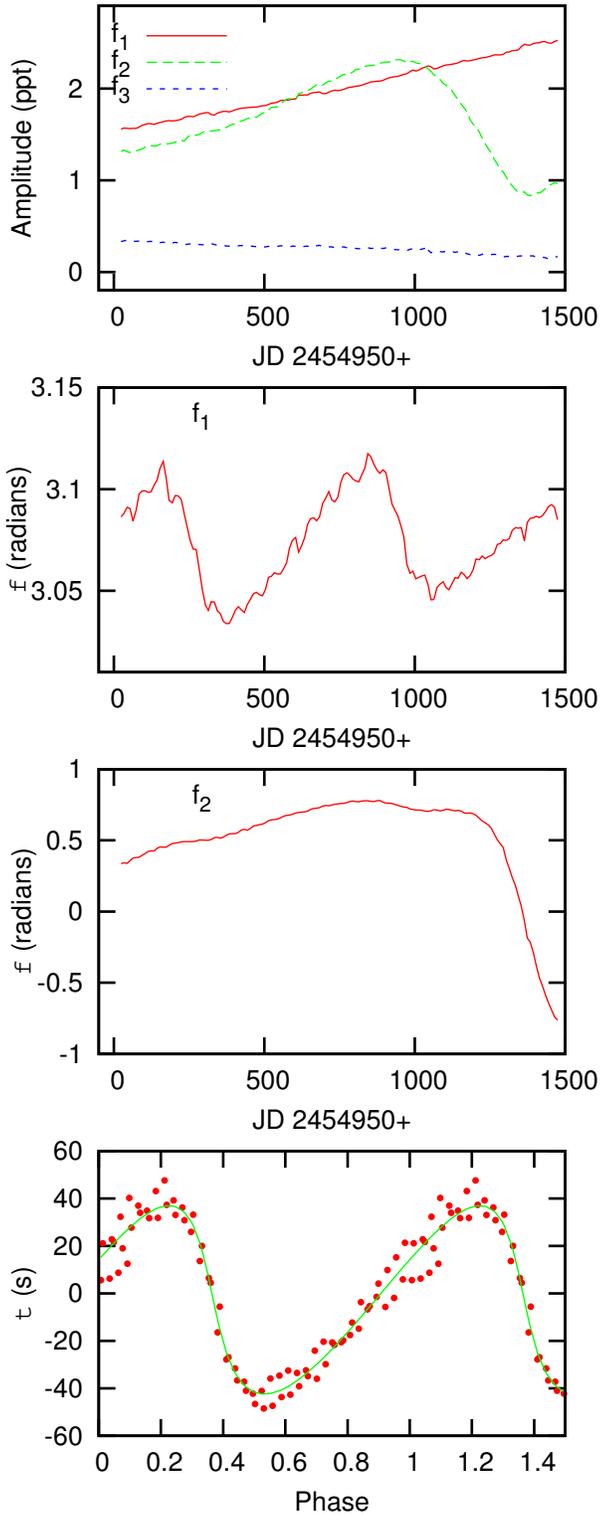} 
\caption{Top panel: amplitude variation of the three highest amplitude 
pulsation modes in KIC\,3429637 from {\it Kepler} photometry. The next
panels from the top shows the phase variation of $f_1$ and $f_2$.
In the bottom panel the time delay variation, $\tau$, derived from $f_1$ 
and $f_2$ are phased with the orbital period of 704.7\,d.  The solid line is
from the orbital solution. The time of phase zero is JD\,2454950.0.}
\label{k003429637} 
\end{figure} 

For KIC\,3429637, \citet{Catanzaro2011} obtained $T_{\rm eff} = 7300 \pm
200$\,K, $\log g = 3.16 \pm 0.25$  and $v \sin i = 50 \pm  5$\,km\,s$^{-1}$, values 
later confirmed by \citet{Murphy2012b}.  The remarkable amplitude increase in 
the $\delta$~Sct pulsations of this star has been discussed by 
\citet{Murphy2012b}.  Since that time, additional {\it Kepler} data show a more
complex amplitude variation for the frequencies of largest amplitude 
(Fig.\,\ref{k003429637}, top panel).  While $f_1 = 10.337$\,d$^{-1}$ and
$f_3 = 10.936$\,d$^{-1}$ show steady amplitude changes, the steady amplitude
increase for $f_2 = 12.472$\,d$^{-1}$ reverses and decreases sharply around
JD\,2455950.  The latest observations indicate that the amplitude of $f_2$
is, once again, increasing.  The timescale seems to be too short for an
evolutionary explanation, as suggested by \citet{Murphy2012b}.

KIC\,5219533 has a  very rich $\delta$~Sct frequency spectrum, including a 
large number of modes in the $\gamma$~Dor range.  The projected rotational
velocity is $v \sin i = 115$\,km\,s$^{-1}$.  The principal peak at
$f_1 = 10.2853$\,d$^{-1}$ is clearly variable in frequency.  The phase, 
$\phi$, varies smoothly in the range $2.0 < \phi < 4.0$\,radians 
over 400\,d which corresponds to approximately 0.0008\,d$^{-1}$.

A recent abundance analysis of KIC\,9117875 by \citet{Catanzaro2014} indicates
$T_{\rm eff} = 7400 \pm 150$\,K, $\log g = 3.6 \pm 0.1$\,dex, and  
$v \sin i = 58 \pm 6$\,km\,s$^{-1}$.  Both Ca and Sc show underabundances  of 
about 1~dex, while the heavy elements are all overabundant: 0.5 dex for 
iron-peak elements and  about 2 dex for Ba.   The Am nature of this star is 
thus confirmed.  The {\it Kepler} light curve shows beating and low frequencies
resembling a $\gamma$~Dor star and it was classified as such by 
\citet{Uytterhoeven2011}.  However, a more detailed examination of the 
periodogram shows a broad feature at about $f_1 = 0.672$\,d$^{-1}$ with a very
sharp neighbouring peak at $f_2 = 0.7157$\,d$^{-1}$.  Several harmonics of 
$f_2$ are visible, which is quite unlike the structure found in typical 
$\gamma$~Dor stars.  In fact, this mysterious structure is found in a 
considerable fraction of A stars, but is unexplained \citep{Balona2013c}.  
\citet{Balona2014b} suggests that the sharp feature might be explained by a 
synchronously orbiting planet, not necessarily transiting.  Assuming that
$f_1$ is the rotation frequency and using the above projected rotational 
velocity leads to a radius $R \sin i \approx 1.69 R_\odot$, which is 
reasonable for an A star.

\begin{figure} 
\centering
\includegraphics[]{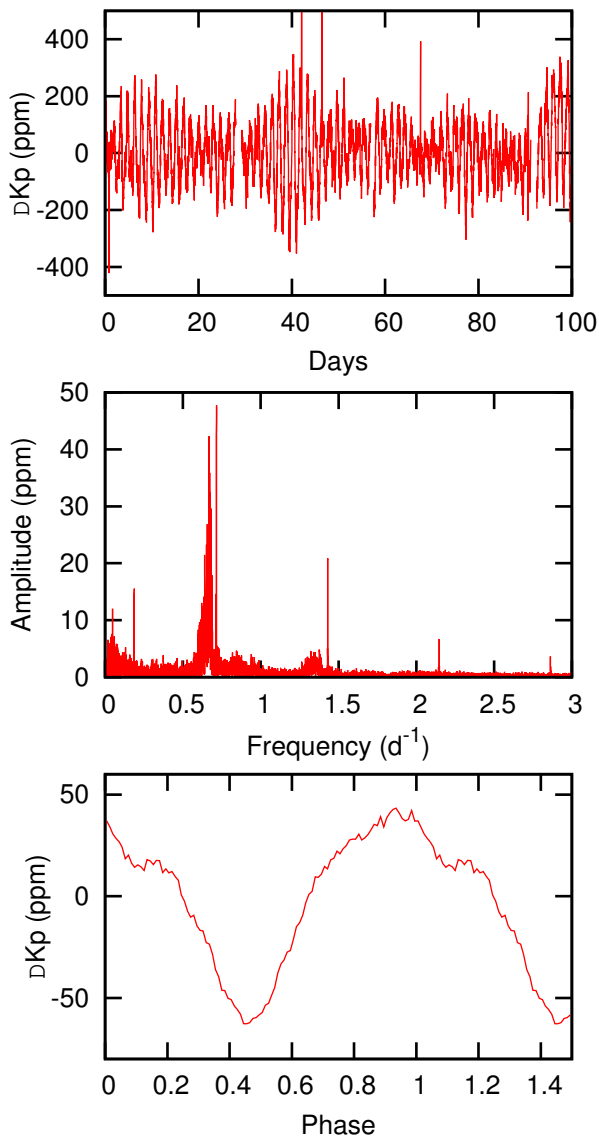} 
\caption{Top panel: part of the {\it Kepler} long-cadence light curve of
KIC\,9117875.  The middle panel is the periodogram showing the broad
feature, $f_1$ and the sharp peak $f_2$ and its many harmonics.  The bottom
panel shows the light curve when all frequencies except for $f_2$ and its
harmonics have been removed.  The data points have been averaged for clarity.}
\label{k009117875} 
\end{figure} 

\begin{figure} 
\centering
\includegraphics[]{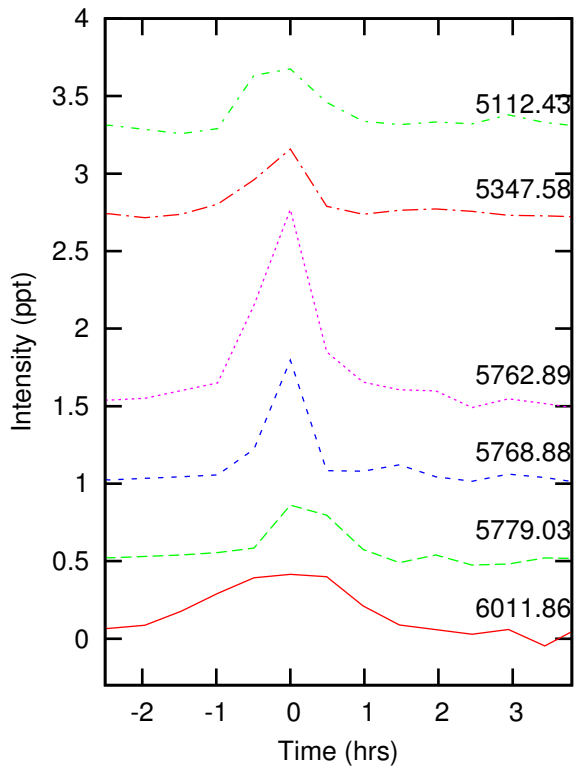} 
\caption{Possible flares in LC data of KIC\,9117875.  The labels give the JD
of peak flare intensity relative to JD\,2450000.00.}
\label{f009117875} 
\end{figure}

There is a further mystery regarding this star in that it appears to flare.
In Fig.\,\ref{f009117875} we show examples of possible flares in this star
in the {\it Kepler} LC data.   It should be noted that LC data is not ideal
to detect flares since the exposure time of 30\,min is a considerable
fraction of the duration of the flare.  Short-cadence data are available for
this star, but only for about 10\,d during Q0.

\begin{figure} 
\centering
\includegraphics[]{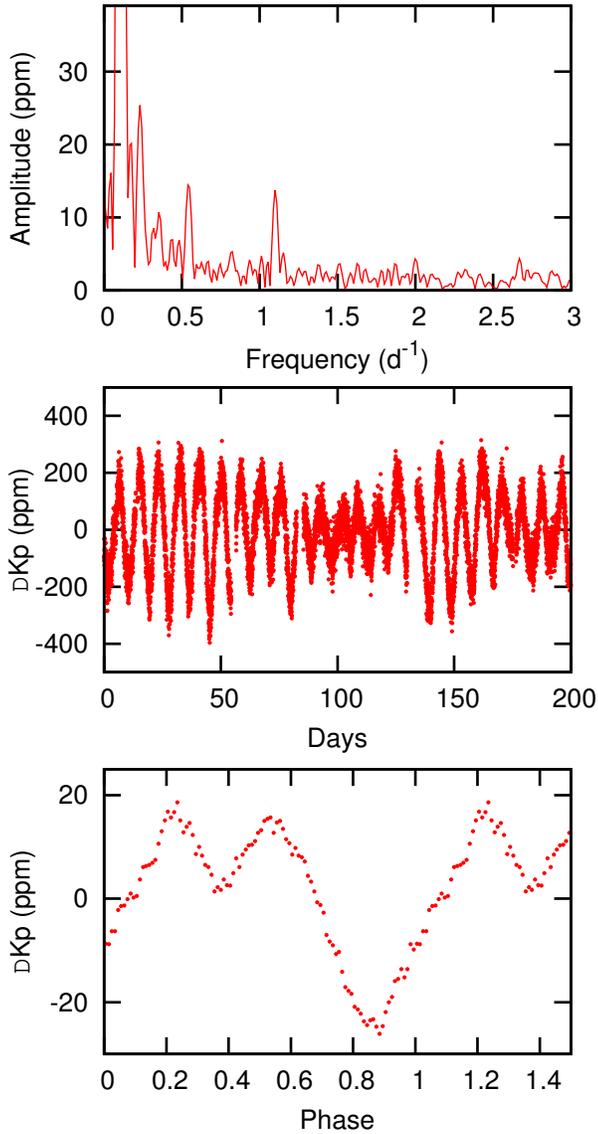} 
\caption{Top panel: periodogram of KIC\,9204718 showing several low
frequencies.  Middle panel: part of the light curve showing the dominant 
0.1145\,d$^{-1}$ frequency.  Bottom panel: phased and averaged light curve
showing the 0.5474\,d$^{-1}$ light variation.}
\label{k009204718}
\end{figure}

KIC\,9204718 was confirmed as an Am star by \citet{Catanzaro2014} who
find $T_{\rm eff} = 7600 \pm 150$\,K,  $\log g = 3.8 \pm 0.1$\,dex and 
$v \sin i = 27 \pm 3$\,km\,s$^{-1}$. There are clearly several low-frequency 
peaks in the periodogram, as shown in Fig.\,\ref{k009204718}.  The largest 
peak is at 0.1145\,d$^{-1}$ and its harmonic.  These peaks are both rather 
broad suggesting amplitude or frequency modulation; in fact, the light 
variation with this frequency is clearly visible in the light curve
(Fig.\,\ref{k009204718}).  This variation is suggestive of a starspot, in 
which case 0.1145\,d$^{-1}$ would be the rotation frequency.  This is 
reasonable since we measured $v \sin i = 27 \pm 3$\,km\,s$^{-1}$.

Even more interesting is the presence of several harmonics of 0.5474\,d$^{-1}$. 
If we remove the 0.1145\,d$^{-1}$ and the $\delta$~Sct frequencies from the
data, what is left is the 0.5474\,d$^{-1}$ variation and its harmonics. 
By binning the light curve and averaging the data in each bin, we obtained
the phased light curve shown in the bottom panel of Fig.\,\ref{k009204718}. 
There is no simple explanation for this variation.  It cannot be the
rotational frequency, since this has already been identified.  The light
curve resembles that of a contact binary, but if so it implies that the
rotational period is not synchronized with the orbital period.

There is a great deal of uncertainty regarding the nature of KIC\,9272082.
\citet{Macrae1952} noted its possible peculiar spectrum, but did not give any 
details.  \citet{Bertaud1960} gives a classification of A5m, but 
\citet{Floquet1970} classified it as a normal A7 star.   The star is included 
as an Am star in the \citet{Renson2009} catalogue.  \citet{Catanzaro2014} 
found it to be an A4 main sequence star with  $T_{\rm eff} = 8500 \pm 200$\,K,
$\log g = 3.9 \pm 0.1$\,dex and $v \sin i = 75 \pm 7$\,km\,s$^{-1}$.  Most of 
the elements in this star are somewhat overabundant by about 0.5--1.8~dex.  
However, Si, Cr, Ni, and Ba have normal abundances.  Although the abundance 
pattern is not typical, it does not conform to that expected for Am stars.

\begin{figure}
\centering
\includegraphics[]{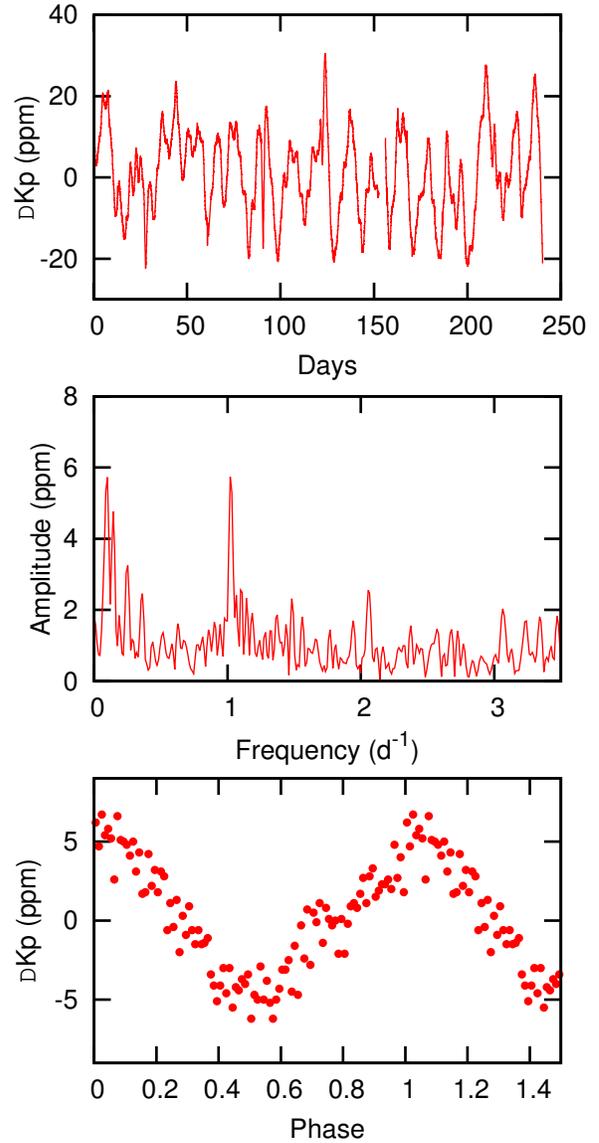} 
\caption{Top panel: portion of the corrected {\it Kepler} light curve of 
KIC\,9272082.  The light curve has been smoothed by a Gaussian filter with
FWHM = 0.5\,d for better visibility.  Middle panel: periodogram of the LC data.
Bottom panel: binned light curve phased with frequency $f_1 = 1.03058$\,d$^{-1}$.}
\label{k009272082}
\end{figure} 

The light curve (top panel of Fig.\,\ref{k009272082}) shows distinct
variations which are confirmed by the periodogram (middle panel).  There is
a constant frequency $f_1 = 1.03058$\,d$^{-1}$ and its harmonic, suggestive
of rotational modulation.  The corrected LC data phased with this frequency
shows a roughly sinusoidal variation (Fig.\,\ref{k009272082}), which we
assume to be rotational modulation.

\section{Binary motion from time delay}

It is generally thought that most Am stars are binaries with orbital periods 
in the range 1--10\,d \citep{Abt1967}.  It is of interest to determine whether
any of the stars in our sample are spectroscopic binaries.  A pulsating star 
in a binary system acts as a clock.  If the star is a member of a binary
system, its distance will vary as it orbits the barycentre.  The changing
distance leads to a variable phase of pulsation.  The method has been
discussed by \citet{Shibahashi2012} using data transformed to the frequency 
domain.  More recently, \citet{Murphy2014} has used the more direct analysis
of phase variation.  Application to $\delta$~Sct stars is difficult because
of the many close frequencies which distort the phase variation.  By
direct minimization of the photometric error, it is possible to construct a
diagram similar to the periodogram where the orbital semi-major axis is
shown as a function of the orbital period (the binarogram).  This technique 
allows relatively quick detection of binary candidates \citep{Balona2014c}.
 
KIC\,3429637 has been discussed above in the context of its $\delta$~Sct
pulsations.  Fig.\,\ref{k003429637} also shows the phase variations of 
$f_1$ and $f_2$.  We used a window with semi-width of 20\,d and a sampling 
interval of 20\,d.  The dominant frequency, $f_1$, shows a clear periodic 
variation with $P = 704.7$\,d.  There is a rather sudden change in phase of 
$f_2$ (corresponding to a change in frequency) at a time soon after the sharp 
decrease in amplitude shown in the top panel.  Close inspection reveals that 
the 704.7-d period is also present in the phase variation of $f_2$.  
Short-cadence observations are available for this star, so there is no 
ambiguity regarding the frequencies.

We applied the time-delay algorithm described above to both $f_1$ and $f_2$.
Because there is a strong sudden decrease of amplitude and frequency around
JD\,2455950, we decided to use only the data up to this date.  There is
still a long-term trend which was removed by fitting a Fourier series
combined with a polynomial of third degree.  The derived time delay variation 
for $f_1$ and $f_2$ are in good agreement and both are shown in the bottom 
panel of Fig.\,\ref{k003429637} phased with the orbital period.  By fitting 
the time delay variation we obtain the orbital solution shown in 
Table\,\ref{taborbit} and shown in the bottom panel of Fig.\,\ref{k003429637}.

\begin{table}
\begin{center}
\label{taborbit}
\caption{Binary elements of KIC\,3429637 from the time delay of the {\it Kepler} 
light curve.  The orbital period $P_{\rm orb}$, half-range of primary radial velocity 
variation $K_1$, eccentricity $e$, longitude of periastron $\omega$, 
JD of periastron $T_{\rm per}$, projected semi-major axis
of the primary $a_1 \sin i$, and mass function  $f(M)$.}
\begin{tabular}{lrl}
\hline
Parameter          &  Value                         & Units       \\
\hline
$P_{\rm orb}$      & $704.7 \pm 4.4$                & days        \\
$K_1$              & $1.79 \pm 0.08$                & km s$^{-1}$ \\
$e$                & $0.61 \pm 0.04$                &             \\
$\omega$           & $3.05 \pm 0.02$                & radians     \\
$T_{\rm per}$      & $2455204.6 \pm 3.6$            &             \\
$a_1 \sin i$       & $0.100 \pm 0.005$              & AU          \\
$f(M)$             & $(2.69 \pm 0.03)\times 10^{-4}$& $M_\odot$   \\
\hline
\end{tabular}
\end{center}
\end{table}

\begin{figure}
\centering
\includegraphics[]{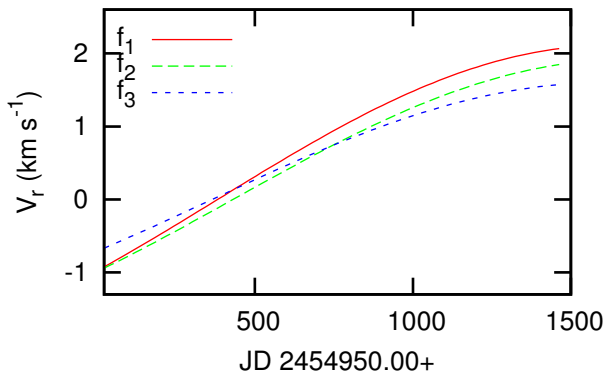} 
\caption{Radial velocity curves derived from frequencies $f_1$, $f_2$ and
$f_3$ in KIC\,11445913.}
\label{k011445913}
\end{figure}

For KIC\,11445913 short cadence observations show that the dominant peaks are 
at $f_1 = 31.5578$, $f_2 = 25.3771$, $f_3 = 22.1330$ and 
$f_4 = 37.8204$\,d$^{-1}$. As usual there are quite a number of low-frequency 
peaks, but none that could be identified as due to starspots.  The value 
$v \sin i = 51 \pm 1$\,km\,s$^{-1}$ \citep{Balona2011d}.  Time delay analysis 
of $f_1$, $f_2$ and $f_3$ shows that they have the same phase variation, but 
the period, if it exists, is considerably longer than the almost 4\,yr 
duration of the {\it Kepler} data.   Fig.\,\ref{k011445913} shows the resulting 
radial velocities derived from the first derivative of the time delay
variation.  This star might be a binary with a period of around 10\,yr or longer.

\section{K2 light curves}

Owing to the failure of two of the four reaction wheels, the {\it Kepler} 
spacecraft has lost its pointing ability.  The telescope drift is corrected
at regular intervals, with the result that there is substantial image drift
and a large increase in photometric error from about 20~ppm for the original 
{\it Kepler} field to about 300\,ppm for the K2 field.  Here we present
results for Am stars observed in the first K2 scientific mission, Campaign
0.  The observations occur over a timespan of about 77\,d, but with a large
gap of 26.5\,d.  The total on-target time is about 47\,d.  This rather short
time and the much increased scatter severely compromises detection of
significant frequencies less than about 3\,d$^{-1}$.

\begin{figure}
\centering
\includegraphics[]{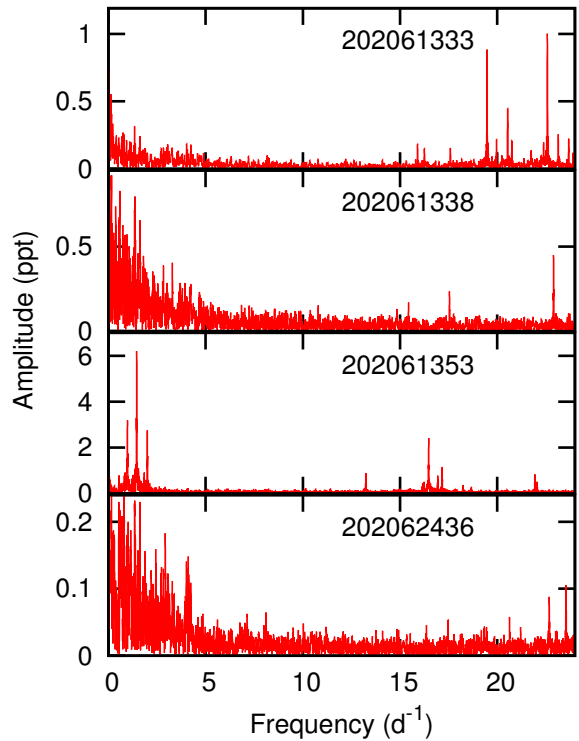} 
\caption{Periodograms of Am $\delta$~Sct stars observed in the K2 Campaign 0.}
\label{k2dsct}
\end{figure}

Many of these stars were observed by the {\it STEREO} spacecraft.  No 
significant variability could be found for EPIC\,202059291, 202059336, 
202061329, 202061333, 202061336, 202061338 and 202061363 from {\it STEREO}
observations \citet{Paunzen2013}.  The remarkably low effective temperature 
for EPIC\,202061363 reported in \citet{Ammons2006} is confirmed by our own
analysis of available multicolour photometry.  {\it STEREO} observations of 
EPIC\,202061349  also do not show any significant variability \citep{Wraight2012}.  
Of these stars, we detected variability in EPIC\,202059291, 202061333 and 
EPIC\,202061338 (see below).  Some of the stars are $\delta$~Sct variables.  
The periodograms of the $\delta$~Sct stars are shown in Fig.\,\ref{k2dsct}.
The effective temperature of EPIC\,202062436 was obtained through our own
analysis of the multicolour photometry.

EPIC\,202061353 is a known $\delta$~Sct star from {\it SuperWasp}
photometry \citep{Smalley2011}.  This is confirmed by the K2 data
(Fig.\,\ref{k2dsct}).  The main feature is a triplet of low-frequency
peaks at 1.452, 0.980 and 2.000\,d$^{-1}$ which are not quite
equally spaced.

\begin{figure}
\centering
\includegraphics[]{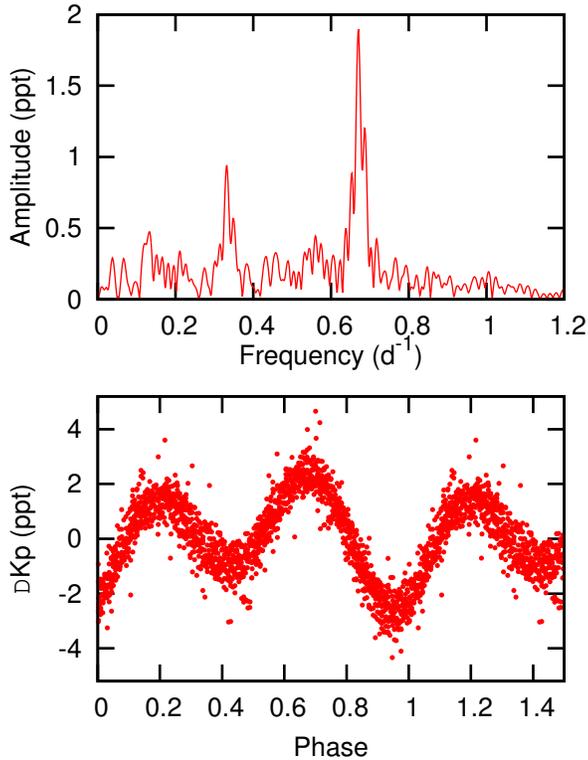} 
\caption{Top panel: periodogram of EPIC\,202059291. Light curve phased with
period $P = 2.984$\,d.}
\label{k202059291}
\end{figure}

The periodogram of EPIC\,202059291 shows only two significant peaks.  
Considering the very low amplitude, the variability is best understood as 
rotational modulation with period $P = 2.984$\,d, but it could also be 
interpreted as a contact binary.  One of us (GC) obtained a single spectrum 
of this star which appears to be a double-lined spectroscopic binary.  It
can probably be ruled out as an Am star.  The periodogram and  phased light 
curve is shown in Fig.\,\ref{k202059291}.

\begin{figure}
\centering
\includegraphics[]{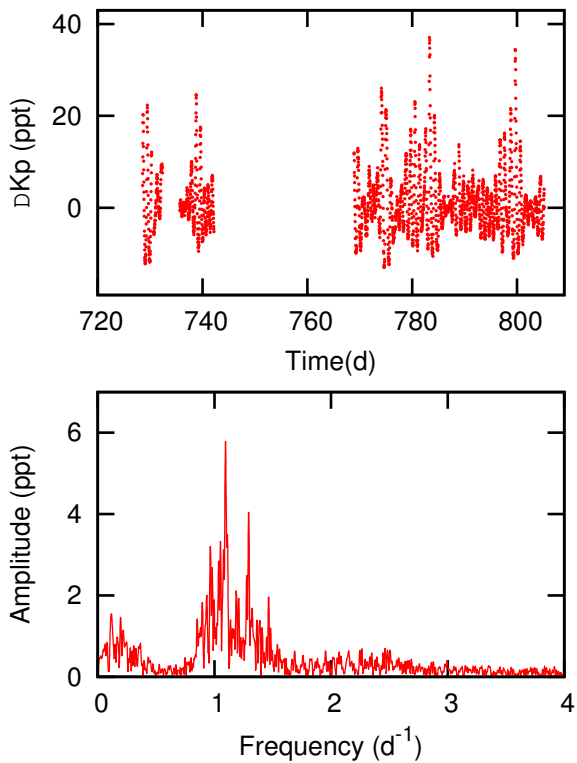} 
\caption{Top panel: light curve of EPIC\,202062447 showing typical asymmetric
light curve of a $\gamma$~Dor star of type ASYM.  The bottom panel shows the
periodogram.}
\label{k202062447}
\end{figure}

The light curve of EPIC\,202062447 (Fig.\,\ref{k202062447}) shows that it 
is a $\gamma$~Dor of the ASYM type \citep{Balona2011f}.  The periodogram 
shows dominant periods of 0.912\,d and 0.773\,d.

The field of EPIC\,202062450 contains two stars of almost equal brightness 
separated by 5\,arcsec.  The brighter star, FI~Ori, is a detached eclipsing 
binary with  $P = 4.4478$\,d and spectral type K0III/IV \citep{Halbedel1985}.  
The eclipses are very easily seen in the K2 combined light of the two stars.
The fainter star, HD\,256320\,B has a spectral type kA5hA8mF5 \citep{Halbedel1985}.  
We constructed an aperture which only included this star, but could not find 
any evidence of periodic variation with an amplitude exceeding 5\,ppt.

\section{Discussion and conclusion}

It is evident from Table \ref{amtab} that most Am stars in the  {\it Kepler} 
field exhibit rotational modulation.  An idea of the spot size may be gained 
from the light amplitude of the modulation shown in Table\,\ref{amtab}.
There are only 10 Am stars in the {\it Kepler} field known to be rotational
variables and the amplitude range is large.  The mean amplitude is 
$93 \pm 25$\,ppm.  The mean rotational amplitude from 875 A stars in 
\citet{Balona2013c} is $541 \pm 86$\,ppm.   This seems to suggest that, in
general, spots on Am stars are smaller than on normal A stars, but it is not
possible to draw a definite conclusion owing to the small sample.    

One or more flares appear to be seen in two confirmed Am stars, KIC\,8703413 
and KIC\,9117875.  The relatively large flare intensities imply that if
the origin of the flare is a cool companion, the flare must be about two
orders of magnitude more intense than on an isolated cool star owing to the
much larger luminosity of the Am star.  While this is not impossible, it
seems very improbable.  It is clear from a study of a much larger number of 
A stars that the flares cannot all be attributed to a cool companion
\citep{Balona2012c, Balona2013c}.

The presence of starspots and flares on Am stars suggests that Am
stars may have significant magnetic fields similar in strength to normal A
stars.  The magnetic fields in both Am and normal A stars are sufficiently 
intense to form starspots and flares.  Hence the idea that the superficial 
metal enrichment in Am stars is a result of the effects of diffusion and 
gravitational settling in the absence of a magnetic field may need to be 
revised.   Furthermore, the relative number of pulsating Am stars is similar 
to the relative number of $\delta$~Sct stars among the A star population.  Am 
stars which are $\delta$~Sct variables occupy the same region of the 
instability strip as normal $\delta$~Sct stars \citep{Smalley2011}.  All these
observations are contrary to the view that Am stars are formed by diffusion of
elements in the absence of a magnetic field and that $\delta$~Sct pulsation in
Am stars only occurs near the red edge of the instability strip.  Given these 
facts, one has to conclude either that the diffusion model for Am stars needs 
to be revised or that it is incorrect.

Thermohaline convection takes place when a layer of enhanced metal abundance
is formed above layers of lighter mean molecular weight.  This happens in the 
case of accretion of predominantly metal-rich material \citep{Vauclair2004}, 
and also when such a layer is formed due to atomic diffusion.   This
important effect has only recently been studied.  \citet{Theado2009b} finds
that accumulation of heavy elements due to atomic diffusion is attenuated
when thermohaline mixing is taken into account, though not completely 
suppressed.   The inclusion of the thermohaline instability in A-star models 
needs to be more fully investigated to see whether this can reconcile the
diffusion model of Am stars with observations.

The presence of pulsations in Am stars is only one symptom of a more general
problem which has been encountered since the advent of {\it Kepler}
photometry.  For example, it is not understood why less than half of stars
in the $\delta$~Sct instability strip pulsate \citep{Balona2011g}, or why
low-frequency pulsations are a general feature of all $\delta$~Sct stars
\citep{Balona2014a}.  Further exploration of pulsational driving needs to be
undertaken.  For example, coherent pulsations may be driven by turbulent 
pressure in the hydrogen ionization zone \citep{Antoci2014}. 

It is well established that stars hosting planets are metal rich, but the
reason for such metal enhancement is still a subject for debate.  The
enhancement may be primordial, the result of accretion or both.  The 
argument against an accretion origin is related to the mass of the outer 
convective zones, which varies by more than one order of magnitude among 
the considered stars, while the observed over-abundances of metals are 
about the same.  \citet{Vauclair2004} found that thermohaline convection
mixes the accreted matter and leaves only a very small $\mu$-gradient at 
the end of the mixing process.  The  remaining $\mu$ gradient is too small 
to account for the observed overmetallicity (see also \citet{Theado2012}).  
It should be noted, however, that all these calculations are confined to 
solar-type stars.  The effect of thermohaline convection on intermediate 
mass A stars has not yet been determined.

Considering the problems encountered by standard diffusion theory in
explaining the properties of Am stars, it may be appropriate to again
consider the possibility that the Am phenomenon may be a combination of
diffusion with accretion. For example, \citet{BohmVitense2006} has argued
that accretion of interstellar material by A stars with tangled magnetic 
fields, which are weaker than those in peculiar A (Ap) stars, has the best 
chance of explaining the main characteristics of the peculiar heavy-element 
abundances in Am star photospheres.
Another look at accretion of close-in planets should also be made.  We know,
for example, that accretion of comet-sized bodies is very likely occurring in 
the A6V star $\beta$~Pic which is surrounded by a debris disk \citep{Beust1996}.
There is no reason to suppose that this star is unique and we can expect
infalling material, even of planets, to occur.  As \citet{Pinsonneault2001} 
has pointed out, the mass of the outer convective zone decreases rapidly for
hot stars, being very thin in A stars.  Therefore one may expect infalling
material, mixed only in the thin outer convective zone, to have a
substantial effect on the observed metallicity.  For example, the mass above
the base of the He\,II convective zone in a typical mid-A star is less than
the mass of the Earth.  Since the original fraction of metals in the star 
is only about 2 percent of the mass of the zone, even a body much smaller 
than the Earth could increase the apparent metal abundance by a factor of 
two or more.  It would be surprising if no such impacts ever occurred.  
Therefore one should not be surprised by the peculiar metal abundances in 
some A stars.

The chemically peculiar $\lambda$~Boo stars have an unusually low superficial 
abundance of iron peak elements.  It is thought that this may be a result of
accretion of interstellar material as the star travels through a diffuse
interstellar cloud \citep{Kamp2008}.  However, the abundance pattern in 
$\lambda$~Boo stars is very different from Am stars.  This may reflect 
differences in abundance between the metal-poor interstellar medium 
and the metal-rich composition of infalling small planetary bodies.

\section*{Acknowledgments} 

This paper includes data collected by the {\it Kepler} mission. Funding for the
{\it Kepler} mission is provided by the NASA Science Mission directorate.
The authors wish to thank the {\it Kepler} team for their generosity in
allowing the data to be released and for their outstanding efforts which have
made these results possible.  

Much of the data presented in this paper were obtained from the
Mikulski Archive for Space Telescopes (MAST). STScI is operated by the
Association of Universities for Research in Astronomy, Inc., under NASA
contract NAS5-26555. Support for MAST for non-HST data is provided by the
NASA Office of Space Science via grant NNX09AF08G and by other grants and
contracts.
 
LAB wishes to thank the South African Astronomical Observatory and the
National Research Foundation for financial support.

\bibliographystyle{mn2e}
\bibliography{amspot}
 
\label{lastpage}

\end{document}